\documentclass[fleqn,usenatbib]{mnras}
\usepackage{ae,aecompl}
\usepackage{graphicx}	% Including figure files
\usepackage{amsmath}	% Advanced maths commands
\usepackage{amssymb}	% Extra maths symbols
\usepackage{color}
\newcommand{\be}{\begin{equation}}
\newcommand{\ee}{\end{equation}}
\newcommand{\cR}{{\cal R}}

\title[]{On the spectrum of stable secondary nuclei in cosmic rays}

\author[]{
P. Blasi,$^{1,2}$\thanks{E-mail: blasi@arcetri.astro.it}
\\
$^{1}$ INAF/Osservatorio Astrofisico di Arcetri, Largo E. Fermi, 5 - 50125 Firenze, Italy\\
$^{2}$ Gran Sasso Science Institute, Viale F. Crispi 7 - 67100 L' Aquila, Italy
}

\date{Accepted XXX. Received YYY; in original form ZZZ}

\pubyear{2017}

\begin{document}
\label{firstpage}
\pagerange{\pageref{firstpage}--\pageref{lastpage}}
\maketitle

\begin{abstract}
The ratio of the fluxes of secondary and primary nuclei in cosmic rays has long been used as an indicator of the grammage traversed in the journey of cosmic ray particles throughout the Galaxy. The basic idea is that primary particles are accelerated in astrophysical sources, such as supernova remnant shocks and eventually propagate in the Galactic volume, occasionally interacting with gas, mainly in the disc of the Galaxy, and there they produce secondary nuclei through spallation. At sufficiently high energy, typically $\gtrsim 100$ GeV/n, the ratio of fluxes of the secondary nucleus to that of the main primary nucleus is found to scale as $E_{k}^{-\delta}$, where $E_{k}$ is the energy per nucleon (a conserved quantity in spallation reactions) and $\delta$ identifies the energy dependence of the diffusion coefficient. The same shock waves that may be responsible for cosmic ray acceleration in the first place also pick up any other charged particle in the upstream, provided being above threshold for injection. The secondary nuclei produced by spallation in the interstellar medium are no exception, hence they also get accelerated. This effect is unavoidable, only its strength may be subject of debate. We compute the spectrum of secondary elements such as boron and lithium taking into account shock reacceleration and compare our predictions with the recent observations of the B/C ratio and preliminary measurements of the boron and lithium flux. Both these sets of data seem to confirm that reacceleration of secondary nuclei indeed plays an important role, thereby affecting the validity of those scaling rules that are often used in cosmic ray physics. 
\end{abstract}

%%%%%%%%%%%%%%%%%%%%%%%%%%%%%%%%%%%%%%%%%%%%%%%%%%

%%%%%%%%%%%%%%%%% BODY OF PAPER %%%%%%%%%%%%%%%%%%
\section{Introduction}
\label{eq:intro}

In cosmic ray (CR) physics, secondary nuclei play a crucial role: the ratio of fluxes of stable secondary nuclei and their parent primary nuclei provides a unique estimate of the grammage traversed by CRs during propagation in the Galaxy. Unstable secondary nuclei, on the other hand, carry information on the escape time of CRs in the Galaxy. The two quantities, escape time and grammage, are clearly connected to each other and their measurement has long been considered as a test of consistency of our picture of diffusive-convective CR transport. Information on the grammage can also be gathered from measurements of the ratio of fluxes of antiprotons. 

The recent measurements carried out with PAMELA and AMS-02 have opened new and unexpected scenarios in our view of the origin of CRs: PAMELA and AMS-02 have discovered that the spectra of protons and helium nuclei are characterized by a break at rigidity $\sim 300$ GV \cite[]{Adriani:2011cu,Aguilar:2015ooa,Aguilar:2015ctt}, which may reflect a new piece of physics involved in either the acceleration or the transport of CRs. A similar break might be present also in the spectra of heavier nuclei \cite[]{Ahn:2010gv} as it also seems to be the case based on preliminary AMS-02 data \cite[]{YanCern2017}. Several possibilities have been put forward to explain this spectral break: some choices of the spatial dependence of the diffusion coefficient would lead to a spectral break \cite[]{Tomassetti:2012ga} at a rigidity that depends on the choice of parameters; the presence of both self-generation of waves and pre-existing turbulence would also naturally lead to a spectral break \cite[]{Blasi:2012yr} at rigidity $\sim 100-1000$ GeV. It has also been proposed that the spectral break may reflect the accidental proximity of a source \cite[]{2012MNRAS.421.1209T}, but this possibility appears rather unnatural in that the probability of occurrence is very low \cite[]{2016arXiv161002010G}. It has also been suggested that the spectral break may reflect the concavity in the spectrum of accelerated particles in the presence of non-linear effects \cite[]{2013ApJ...763...47P}. 

The secondary to primary ratios are powerful tools to discriminate among these possibilities: models that explain the spectral break in terms of acceleration are not expected to affect the primary to secondary ratios, while models based on modifications of CR propagation in general do. On the other hand, if the break reflects the accidental presence of a nearby source, then the flux of secondary nuclei should remain unaffected because it is dominated by the propagation of CRs on large scales. 

In the standard approach to CR transport, at high enough energy (typically $\gtrsim 100$ GeV) the secondary to primary ratio drops with energy proportional to the grammage traversed by CRs, which in turn scales as the diffusion coefficient. If the effective diffusion coefficient has a break in its energy dependence then the ratio manifests the same break. In fact, even the model of \cite{Tomassetti:2012ga}, where the diffusion coefficient has a different energy dependence in the region close to and far from the Galactic disc, can be imagined as CRs experiencing an effective energy dependent diffusion coefficient. Such a model and the one based on CR induced non-linear effects \cite[]{Blasi:2012yr} can be distinguished by measuring the spatial distribution of CRs in the halo, which for obvious reasons is not an easy task.

The simple picture of the secondary to primary ratios reflecting the diffusion properties of CRs is affected by several caveats: for instance the grammage that CRs accumulate inside the sources must appear at sufficiently high energies in the form of a flattening or even a rise \cite[]{2009PhRvL.103e1104B,2009PhRvL.103h1104M}. The strength of this effect depends on whether the medium in which CRs are accelerated is dense or dilute: if core collapse SNe exploding in hot dilute regions are the main contributors to the CR flux it is likely that the grammage accumulated inside the sources is relatively small. If type Ia SNe also accelerate CRs, then one can estimate a grammage of $\sim 0.15~\rm g~cm^{-2}$ as due to the sources \cite[]{Aloisio:2015rsa}, which starts to be visible in the B/C ratio at energies above a few hundred GeV/n. 

The B/C ratio has been recently measured by PAMELA \cite[]{Adriani:2014xoa} and extended to higher energies and with smaller uncertainties by AMS-02 \cite[]{2016PhRvL.117w1102A}. These data were used by \cite{2017arXiv170609812G} to conclude that there is evidence that the spectral break in the primary nuclei is due to diffusion rather than sources. The AMS-02 measurement suggests that the B/C ratio drops with energy as $E_{k}^{-1/3}$ at high energies, in perfect agreement with the naive expectation based on a Kolmogorov spectrum of Galactic turbulence, although this conclusion depends somewhat on the threshold energy above which the fit is calculated. On the other hand, preliminary data from AMS-02 also suggest that the spectrum of lithium, a secondary nucleus that in many respects is expected to behave as boron, shows a pronounced hardening at rigidity above a few hundred GV \cite[]{lithium,YanCern2017}, with a high energy slope of the spectrum that appears to be very similar to that of primary nuclei, in apparent contradiction with the secondary nature of lithium. Since these measurements are going to be used in the near future to test different models of CR propagation, it seems timely to make an assessment of the different reasons why the naive expectation for secondary nuclei might need to be revisited. 

The main reason for expecting a deviation of the spectrum of secondary nuclei from their standard trend is that all nuclei are subject to re-energization at the same shock waves that accelerate their parent nuclei (primaries), and this phenomenon leads to a flattening of the spectrum of secondaries to the same spectrum of primary nuclei, above some critical energy. The effect of re-acceleration at shocks was first discussed by \cite{1987ApJ...316..676W} in the context of a leaky-box-like model of CR transport. The authors found that this phenomenon has negligible effect on the spectrum of primary nuclei but, as discussed below, is crucial for secondary nuclei, since their spectrum is steeper than that of primaries due to diffusive transport in the Galaxy.  

The main goal of the present paper is that of finding a solution of the diffusion equation at shocks and then in the Galaxy that describes the phenomenon of reacceleration of secondary nuclei in a self-consistent manner. The results of this calculation will be compared with the recent measurements of the B/C ratio and of the flux of boron and lithium, which will show evidence that this phenomenon is at work. 

The paper is organized as follows: In \S \ref{sec:reac} we will describe the phenomenon of reacceleration at a shock front and specialize it to the case of secondary nuclei present in the environment where a supernova explodes. In \S \ref{sec:transport} we will derive the solution of the transport equation of CRs in the Galaxy in the presence of reacceleration of secondary nuclei. The results of this calculation will then be compared with data in \S \ref{sec:data}. We will illustrate our conclusions in \S \ref{sec:conclude}.

\section{Acceleration and reacceleration at a shock}
\label{sec:reac}

There are several valuable approaches to diffusive shock acceleration (DSA), but the one that perhaps best describes acceleration of both particles injected at the shock and seed particles is the one based on solving the diffusion-advection equation at the shock: 
\be
\frac{\partial}{\partial z}
\left[ D(p)  \frac{\partial}{\partial z} f(z,p) \right] - 
u  \frac{\partial f (z,p)}{\partial z} + 
\frac{1}{3} \left(\frac{d u}{d z}\right)
~p~\frac{\partial f(z,p)}{\partial p} + Q(z,p) = 0,
\label{eq:trans}
\ee
where $f(z,p)$ is the distribution function of accelerated particles as a function of location $z$ and momentum $p$, $u$ is the fluid velocity, $D(p)$ is the diffusion coefficient, that for simplicity we assume to be independent of $z$, and $Q$ is a function describing injection. The $z$-axis is assumed to go from upstream infinity ($z=-\infty$) to downstream infinity ($z=+\infty$). This stationary equation in one spatial dimension catches the main physical ingredients of the problem of diffusion. The injection of particles at a given momentum $p_{inj}$ at the shock surface is modelled by assuming that:
\be
Q(z,p) = \frac{\eta n_{1} u_1}{4\pi p_{inj}^2} \delta(p-p_{inj})\delta(z) = q_{0}(p) \delta(z),
\label{eq:inj}
\ee
where $\eta$ is an acceleration efficiency in number. Hereafter the index ``1'' (``2'') is used to describe quantities upstream (downstream). For instance $n_{1}$ and $u_{1}$ in equation (\ref{eq:inj}) are the values of gas density and velocity upstream of the shock. Non-linear effects induce the formation of a precursor upstream of the shock (see \cite{2013A&ARv..21...70B} and references therein) but here we will not discuss such effects.  

Following \cite{Bell:1978zc}, the acceleration of pre-existing seed particles (that usually is referred to as {\it reacceleration}) is introduced in the problem by adopting the boundary condition that the distribution function $f$ equals the distribution function of seeds, $g(p)$, for $z\to -\infty$. The solution method of equation (\ref{eq:trans}) is well established: integrating such an equation between $z=0^{-}$ and $z=0^{+}$ one gets:
\be
\left[ D(p)  \frac{\partial f}{\partial z}\right]_{2} - \left[ D(p)  \frac{\partial f}{\partial z}\right]_{1}
+ \frac{1}{3} (u_{2}-u_{1})~p~\frac{\partial f_{0}(p)}{\partial p} + q_{0}(p) = 0,
\ee
where $f_{0}(p)$ is the distribution function of accelerated particles at the shock location, $z=0$. Imposing homogeneity downstream implies that $\left[\frac{\partial f}{\partial z}\right]_{2}=0$. In the upstream region ($z<0$), equation (\ref{eq:trans}) simplifies to
\be
\frac{\partial}{\partial z}
\left[ D(p)  \frac{\partial}{\partial z} f(z,p) - u f \right] = 0,
\ee 
which can be easily solved with the boundary condition that $f(z=-\infty,p)=g(p)$. The particle distribution function at the shock, that takes into account both acceleration of injected particles and reacceleration of seed particles is then easily derived and reads:
$$
f_{0}(p) = \frac{s \eta n_{gas}}{4 \pi p_{inj}^{3}} \left( \frac{p}{p_{inj}} \right)^{-s} +
$$
\be
~~~~~~~~~~+ s  \int_{p_{0}}^{p} \frac{dp'}{p'}  \left( \frac{p'}{p} \right)^{s} g (p'),
\label{eq:f0}
\ee
where we introduced the shock velocity $v_{sh}=u_{1}$. Here, as usual, $s=3r/(r-1)$ with $r=u_{1}/u_{2}$ (compression factor at the shock). We introduced a momentum $p_{0}$ representing the minimum momentum of seed particles: such momentum may or may not be the same as $p_{inj}$. In any case, for the spectra of seed particles considered in this manuscript the choice of $p_{0}$ has no practical implications, provided is low enough (below $\sim$ GeV), because the integral in equation (\ref{eq:f0}) is typically dominated by the upper integration limit. For a strong shock, one has that $r\to 4$ and $s\to 4$. It is worth recalling that whenever the spectrum of seeds is steeper than $\sim p^{-s}$, the spectrum of reaccelerated particles asymptotically approaches $\sim p^{-s}$. For the shocks we are interested in, $s\sim 4$. The spectra of seeds we consider (spectra of primary and secondary nuclei in the Galaxy) are always steeper than $p^{-4}$, at least at energies $\gtrsim 10$ GeV/n. This implies that the second term in equation (\ref{eq:f0}) always returns a contribution to $f_{0}$ that is close to $\sim p^{-s}$. For primary nuclei, this contribution is expected to affect mainly the normalization but not the spectrum. On the other hand, for secondary nuclei the first term vanishes and the second term again returns a contribution that is close to $\sim p^{-s}$. Since the spectrum of secondary nuclei in the Galaxy at high enough energies is $\propto E^{-\gamma-\delta}$ (where $\delta$ defines the energy dependence of the diffusion coefficient), it is clear that the effect of reaccelerated secondary nuclei is bound to become dominant above some critical energy, that will be estimated in the next section.

\section{Galactic transport of cosmic rays in the presence of reacceleration}
\label{sec:transport}

In the assumption that the sources are all located in a thin disc with half-thickness $h_{d}$ where the gas, with density $n_{d}$, is also embedded, the stationary transport equation in one spatial dimension for nuclei of type $\alpha$ reads:
$$
-\frac{\partial}{\partial z} \left[D_{\alpha}(p) \frac{\partial F_{\alpha}}{\partial z}\right] + 2 h_{d} n_{d} v(p) \sigma_{\alpha} \delta(z) F_{\alpha} -
2 h_d q_{0,\alpha}(p) \delta(z) =
$$
\be
 = 
\sum_{\alpha'>\alpha} 2 h_{d} n_{d}\, v(p') \sigma_{\alpha'\to\alpha}\delta(z) F_{\alpha'} (p') \left(\frac{p'}{p}\right)^{2} \frac{dp'}{dp},
\label{eq:trans0}
\ee
where $D_{\alpha}(p)$ is the diffusion coefficient, $F_{\alpha}(p,z)$ is the particle distribution function of nuclei of type $\alpha$. The cross sections $\sigma_{\alpha}$ and $\sigma_{\alpha'\to\alpha}$ refer to the cross sections for spallation of the element $\alpha$ and the cross section for spallation of the nucleus $\alpha'$ to a nucleus $\alpha$. The source term and the spallation terms in equation (\ref{eq:trans0}) are written in the assumption that both take place in an infinitely thin region. This assumption holds for as long as the thickness of the disc is much smaller than the size of the halo and of the range where spallation losses become dominant. In other words: $h_{d}\ll (D(p)\tau_{sp})^{1/2}$, where $\tau_{sp}$ is the time scale for spallation reactions. For the situations of interest for us this condition is always satisfied. Notice that in the last term of equation (\ref{eq:trans0}) we took into account that in the spallation reaction a nucleus of type $\alpha$ with momentum $p$ is produced by a nucleus of type $\alpha'$ with momentum $p'$, chosen in such a way that $p$ and $p'$ correspond to the same kinetic energy per nucleon, namely $p'=(A'/A) p$, where $A$ and $A'$ are the two mass numbers. 

The injection term in equation (\ref{eq:trans0}) can be easily connected with the distribution function $f_{0}$ of accelerated particles in \S \ref{sec:reac}:
\be
q_{0,\alpha}(p) = \frac{f_{0,\alpha}(p) V_{SN}\cR_{SN}}{\pi R_{d}^{2} 2 h_{d}},
\label{eq:q0}
\ee 
where we have implicitly assumed that the sites where CR acceleration takes place are the shocks of supernova remnants and SNe explode at a rate $\cR_{SN}$. Here $V_{SN}$ is the total volume of a SNR filled with energetic particles and $R_{d}$ is the radius of the disc of the Galaxy. For simplicity here we assumed that the injection is homogeneous across the disc of the Galaxy. The volume $V_{SN}$ is a parameter of the problem: it is clear that this setup is not necessarily very realistic but it is not easy to go beyond it, since in principle one should follow the time dependence of the acceleration process and of the SN evolution, that are both rather difficult to model. For the purpose of illustrating the importance of reacceleration, the parameter $V_{SN}$ is meaningful because it regulates the probability for a CR particle to re-cross a SN shock and be re-energized. 

It is useful, following \cite{WeightedSlab}, to introduce, for each nucleus of type $\alpha$, the flux as a function of the kinetic energy per nucleon $E_{k}$: $I(E_{k})dE_{k} = p^{2} v(p) F(p) dp$, where $v(p)$ is the velocity of the nucleus. It can be easily shown that:
\be
I_{\alpha}(E_{k}) = A p^{2} F_{\alpha}(p).
\ee
Using this transformation in equation (\ref{eq:trans0}), we obtain the following equation for $I_{\alpha}(E_{k})$:
$$
-\frac{\partial}{\partial z} \left[D_{\alpha}\frac{\partial I_{\alpha}(E_{k})}{\partial z}\right] + 2 h_{d} n_{d} v(E_{k}) \sigma_{\alpha} \delta(z) I_{\alpha}(E_{k}) = 
$$
\be
 = 2 A p^{2} h_d q_{0,\alpha}(p) \delta(z) + \sum_{\alpha'>\alpha} 2 h_{d} n_{d}\, v(E_{k}) \sigma_{\alpha'\to\alpha}\delta(z) I_{\alpha'} (E_{k}),
\label{eq:trans1}
\ee
where we used explicitly the fact that spallation reactions conserve kinetic energy per nucleon. The injection term $q_{0,\alpha}(p)$ is calculated at $p=A\sqrt{E_{k}^{2}+2m_{p} E_{k}}$ and, based on the discussion in \S \ref{sec:reac}, is made, in general, of two contributions: nuclei of type $\alpha$ freshly accelerated at the shock and nuclei of type $\alpha$ already present in the environment and eventually reaccelerated. For secondary nuclei, such as boron and lithium, only the latter contribution to injection is present. In the following we discuss the case of primary and secondary nuclei separately. 

\subsection{The case of primary nuclei}
\label{sec:primary}

For primary nuclei, such as carbon and oxygen, the contribution coming from spallation of heavier elements is negligible and one can write equation (\ref{eq:trans1}) as
$$
-\frac{\partial}{\partial z} \left[D_{\alpha}\frac{\partial I_{\alpha}(E_{k},z)}{\partial z}\right] + 2 h_{d} n_{d} v(E_{k}) \sigma_{\alpha} \delta(z) I_{\alpha}(E_{k}) = 
$$
\be
~~~~~~~~2 A p^{2} h_d q_{0,\alpha}(p) \delta(z) .
\label{eq:primary}
\ee
In order to simplify the notation, we introduce the quantity $\Delta_{V} = V_{SN}/\pi R_{d}^{2} 2 h_{d}$, which represents the ratio of volumes of a typical SNR to the volume of the Galactic disc (typically $\Delta_{V}\sim 10^{-8}$). From equations (\ref{eq:f0}) and (\ref{eq:q0}) follows that 
$$
A p^{2} q_{0,\alpha}(p) = s \Delta_{V} \cR_{SN} \times
$$
\be
\times  \left[ K ~ A \left( \frac{p}{p_{inj}} \right)^{2-s} +  \int_{p_{0}}^{p} \frac{dp'}{p'}  \left( \frac{p'}{p} \right)^{s-2} I_{\alpha}(E'_{k}) \right],
\ee
where $K=\frac{\eta n_{gas}}{4 \pi p_{inj}}$ and the momentum is related to the kinetic energy per nucleon through the relation $p=A\sqrt{E_{k}^{2} + 2 m_{p} E_{k}}$. The important thing to notice is that the function that we wish to solve equation (\ref{eq:primary}) for also enters the injection term (reacceleration). More precisely the reacceleration term is related to the value of the flux $I_{\alpha}(E_{k},z)$ in the disc ($z=0$). Hence equation (\ref{eq:primary}) is best solved by iterations.

For $z\neq 0$ the equation is trivial and under the boundary condition that $I_{\alpha}(E_{k},z=\pm H)=0$ one finds 
\be
I_{\alpha}(z,p) = I_{\alpha,0}(E_{k})\left[ 1 - \frac{|z|}{H}\right],
\label{eq:der}
\ee
where $I_{\alpha,0}(E_{k})=I_{\alpha}(E_{k},z=0)$. On the other hand, integrating equation (\ref{eq:primary}) between $z=0^{-}$ and $z=0^{+}$, one gets:
$$
\left[D_{\alpha}\frac{\partial I_{\alpha}(E_{k},z)}{\partial z}\right]_{0^{+}} = h_{d} n_{d} v(E_{k}) \sigma_{\alpha} I_{\alpha,0} -
$$
\be 
h_{d} s \Delta_{V} \cR_{SN} 
\left[ K  A \left( \frac{p}{p_{inj}} \right)^{2-s} +  \int_{p_{0}}^{p} \frac{dp'}{p'}  \left( \frac{p'}{p} \right)^{s-2} I_{\alpha,0}(E'_{k}) \right].
\label{eq:I0}
\ee
From equation (\ref{eq:der}) one sees that $\left[\frac{\partial I_{\alpha}(E_{k},z)}{\partial z}\right]_{0^{+}}=-I_{\alpha,0}/H$, hence equation (\ref{eq:I0}) leads to:
$$
I_{\alpha,0}^{(i)}(E_{k}) = s \frac{V_{SN}\cR_{SN}}{2\pi R_{d}^{2}H}\frac{H^{2}}{D_{\alpha}} \frac{1}{1+\frac{X(E_{k})}{X_{cr}}}
\times
$$
\be
\times \left[ K  A  \left( \frac{p}{p_{inj}} \right)^{2-s} +  \int_{p_{0}}^{p} \frac{dp'}{p'}  \left( \frac{p'}{p} \right)^{s-2} I_{\alpha,0}^{(i-1)}(E'_{k}) \right],
\label{eq:iterapr}
\ee
where we introduced the grammage:
\be
X(E_{k}) = n_{d}\frac{h_{d}}{H} m v \frac{H^{2}}{D_{\alpha}},
\label{eq:grammage}
\ee
as well as the critical grammage $X_{cr,\alpha}=m/\sigma_{\alpha}$, where $m$ is the mean mass of the interstellar medium gas that acts as target for spallation (we assume $m=1.4 m_{p}$). The quantity $n_{d}h_{d}/H$ that appears in the grammage plays the role of mean density traversed by CRs during propagation in the disc and halo of the Galaxy. The index $(i)$ in equation (\ref{eq:iterapr}) labels the iteration cycle. 

One can estimate the effect of reacceleration on the spectrum of primary nuclei by calculating the result of the first iteration in equation (\ref{eq:iterapr}), namely by taking 
\be
I_{\alpha,0}^{(0)}(E_{k}) = s K A \frac{V_{SN}\cR_{SN}}{2\pi R_{d}^{2}H}\frac{H^{2}}{D_{\alpha}} 
\left( \frac{p}{p_{inj}} \right)^{2-s} \frac{1}{1+\frac{X(E_{k})}{X_{cr}}}
\ee
and replacing it in the integral of equation (\ref{eq:iterapr}), to get:
$$
I_{\alpha,0}^{(1)}(E_{k}) = s K A \frac{V_{SN}\cR_{SN}}{2\pi R_{d}^{2}H}\frac{H^{2}}{D_{\alpha}}
\left( \frac{p}{p_{inj}} \right)^{2-s}\times
$$
\be
\times \left\{
1 + \frac{s V_{SN} \cR_{SN} H^{2}}{2\pi R_{d}^{2} H} \int_{p_{0}}^{p} \frac{dp'}{p'} \frac{1}{D_{\alpha}},
\right\}
\ee
where, for simplicity, we assumed that, at the energies we are interested in, the role of spallation is weak, namely $X(E_{k})\ll X_{cr}$. 

We assume, as it is often done, that the diffusion coefficient is in the form $D_{\alpha}=D_{0}(p/p_{*})^{\delta}$, with $\delta=0$ for $p<p_{*}$, with $p_{*}$ typically in the range of $3-10$ GeV/c. For $p\gg p_{*}$ one finds that the term in parenthesis is 
\be
1 + \frac{s V_{SN} \cR_{SN} H^{2}}{2\pi R_{d}^{2} H D_{0}} \left\{
\ln \left(\frac{p_{*}}{p_{0}}\right) +\frac{1}{\delta}
\right\}.
\label{eq:correction}
\ee
A SN at the beginning of the Sedov phase has a radius of roughly $R_{SN}\sim 2$ pc. For $H\sim 4$ kpc, $\cR_{SN}=1/30~\rm yr^{-1}$, $R_{d}=10$ kpc and $D_{0}=2\times 10^{28}~\rm cm^{2}/s$, one has that the second term in equation \ref{eq:correction} is $4\times 10^{-3}$. In other words, for a young SNR, the role of CRs reaccelerated from the diffuse background is expected to be totally negligible. However, since the volume of a SNR scales as $R_{SN}^{3}$, a radius of a SN of $10$ pc, more suitable for an aged SNR, well inside the Sedov phase, would make this correction of order unity. At high enough energy ($p>p_{*}\sim 10$ GeV/c) the correction due to shock reacceleration becomes independent of energy and one can consider its effect as a correction of order unity to the overall normalization of the flux of primary nuclei. For this reason, we do not explicitly include reacceleration of primary nuclei and reabsorb its effect in the overall normalization of the primary spectra.

\cite{thoudam} investigated a substantially different case, namely the possibility that a population of very weak (typical Mach number $\sim 1.5$) supernova shocks may reaccelerate CRs. When the slope $s$ in the reacceleration term is larger than the slope of the Galactic CR spectrum (say $\sim 4.7$), the reacceleration term does not change the spectrum but only the normalization of the flux. After transport in the Galaxy, according with \cite{thoudam}, the reaccelerated component may become dominant at low energy. However, in order for this effect to be present the weak shocks must 1) have $s\sim 6$ (instead of the standard $s=4$ for strong shocks) and 2) yet be able to accelerate particles to maximum energies in excess of $\sim TeV$. Their result depends critically on the size of these weak shocks, assumed to be $\sim 100$ pc. For instance the effect disappears if a size of 50 pc is assumed. It should be noted that from observations it seems that supernova shocks stop being particle accelerators (their radio emission disappears) when their velocity drops below $\sim 300$ km/s \cite[]{2010A&A...509A..34B} (much higher than the weak shocks invoked by \cite{thoudam}). For this reason here we no longer consider this possibility in the following and we focus instead on reacceleration at the same shocks that are believed to be responsible for the acceleration of the bulk of CRs. 

\subsection{The case of secondary nuclei}

The role of reacceleration is much more prominent on secondary products of hadronic interactions than on primary nuclei. In this section we illustrate this effect on secondary nuclei such as boron and lithium. For the sake of simplicity we limit ourselves to the production of these secondary products in spallation reactions initiated by carbon and oxygen nuclei, whose fluxes will be denoted as $I_{C}(E_{k})$ and $I_{Ox}(E_{k})$. Secondary nuclei are not accelerated from the thermal pool at supernova shocks, hence the direct injection term in equation (\ref{eq:trans1}) vanishes. The transport equation for boron nuclei can be written as follows: 
$$
-\frac{\partial}{\partial z} \left[D_{\alpha}\frac{\partial I_{B}}{\partial z}\right] + 2 h_{d} n_{d} v \sigma_{B} \delta(z) I_{B} = 
$$
$$
\frac{s \cR_{SN} V_{SN}}{\pi R_{d}^{2}} \left( \frac{p}{p_{0}} \right)^{2-s} \int_{p_{0}}^{p} \frac{dp'}{p'}  \left( \frac{p'}{p_{0}} \right)^{s-2} I_{B}(E'_{k})\delta(z)+ 
$$
\be
+ 2 h_{d} n_{d}\, \sigma_{CB}v I_{C} \delta(z) +
2 h_{d} n_{d}\, \sigma_{OxB}v I_{Ox} \delta(z),
\label{eq:transSec}
\ee
where all fluxes are calculated at the same kinetic energy per nucleon $E_{k}$. The flux $I_{B}$ inside the integral is calculated at kinetic energy per nucleon $E'_{k}$ corresponding to the momentum $p'$. The quantity $\sigma_{B}$ is the cross section for spallation of boron nuclei, assumed here to be independent of energy for simplicity, while $\sigma_{CB}$ and $\sigma_{OxB}$ are the cross sections of production of boron from spallation of carbon and oxygen nuclei respectively. In the following we adopt a simplified structure for these cross sections: we parametrize the cross section for spallation of a nucleus of mass $A$ as $\sigma_{A}=45 A^{0.7}~\rm mb$ \cite[]{1984ApJS...56..369L} and we write the cross section for production of a nucleus $A'$ as $\sigma_{AA'}=\sigma_{A}b_{AA'}$, where $b_{AA'}$ is the probability that spallation of the nucleus $A$ leads to production of the nucleus of mass $A'$. For production of boron one has $b_{CB}=0.28$ and  $b_{OxB}=0.11$ \cite[]{Ginzburg:1990sk}. 

Since all terms of production and destruction of boron are localized at $z=0$ (Galactic disc), the spatial dependence of the solution is still in the same form as in equation (\ref{eq:der}), hence after integration between $z=0^{-}$ and $z=0^{+}$ equation (\ref{eq:transSec}) leads to the following expression for the flux of boron:
$$
I_{B,0}(E_{k}) = \frac{I_{C,0}(E_{k})\frac{X(E_{k})}{X_{cr,CB}}}{1+\frac{X(E_{k})}{X_{cr,B}}} +
\frac{I_{Ox,0}(E_{k})\frac{X(E_{k})}{X_{cr,OxB}}}{1+\frac{X(E_{k})}{X_{cr,B}}}+
$$
\be
+ \frac{s\cR_{SN}V_{SN}}{2\pi R_{d}^{2}H}\frac{H^{2}}{D_{B}}\frac{1}{1+\frac{X(E_{k})}{X_{cr,B}}}
\int_{p_{0}}^{p} \frac{dp'}{p'} \left( \frac{p'}{p}\right)^{s-2} I_{B,0}(E'_{k})
\label{eq:reacSec}
\ee
In addition to the grammage (equation \ref{eq:grammage}), here we introduced the critical grammages $X_{cr,B}=m/\sigma_{B}$, $X_{cr,CB}=m/\sigma_{CB}$ and $X_{cr,OxB}=m/\sigma_{OxB}$. In the absence of reacceleration one can see from equation (\ref{eq:reacSec}) that the B/C ratio reads:
\be
\frac{I_{B.0}(E_{k})}{I_{C,0}(E_{k})} = \frac{\frac{X(E_{k})}{X_{cr,CB}}}{1+\frac{X(E_{k})}{X_{cr,B}}} +
\frac{I_{Ox,0}(E_{k})}{I_{C,0}(E_{k})}\frac{\frac{X(E_{k})}{X_{cr,OxB}}}{1+\frac{X(E_{k})}{X_{cr,B}}}.
\ee
In the assumption that the spectra of carbon and oxygen nuclei are the same at high energy, the ratio scales as $\sim X(E_{k})$ provided spallation does not change appreciably the spectrum of any of the species involved, which is expected to be the case at high energies. 

The physical meaning of the reacceleration term is easy to understand: in the absence of this term the high energy spectrum of boron is $I_{B,0}\propto E_{k}^{-s+2-2\delta}$ where $\delta$ refers to the slope of the diffusion coefficient. Replacing such trend in the reacceleration term, one can easily see that the spectrum resulting from reacceleration at an individual SNR is $~E_{k}^{-s+2}$ and after propagation becomes $E_{k}^{-s+2-\delta}$. It follows that there is always a critical energy above which the contribution of reacceleration dominates upon the standard boron flux. In fact, as we discuss below, this contribution is likely to become important (yet not dominant) even below such critical energy.

As we discuss later, recent observations show a rather intriguing situation for lithium nuclei. Hence we also apply the calculations above to the case of lithium as secondary nucleus. The solution of the transport equation is very similar to the one for boron nuclei:
$$
I_{Li,0}(E_{k}) = \frac{I_{C,0}(E_{k})\frac{X(E_{k})}{X_{cr,CLi}}}{1+\frac{X(E_{k})}{X_{cr,Li}}} +
\frac{I_{Ox,0}(E_{k})\frac{X(E_{k})}{X_{cr,OxLi}}}{1+\frac{X(E_{k})}{X_{cr,Li}}}+
$$
\be
+ \frac{s\cR_{SN}V_{SN}}{2\pi R_{d}^{2}H}\frac{H^{2}}{D_{Li}}\frac{1}{1+\frac{X(E_{k})}{X_{cr,Li}}}
\int_{p_{0}}^{p} \frac{dp'}{p'} \left( \frac{p'}{p}\right)^{s-2} I_{Li,0}(E'_{k}),
\label{eq:lithium}
\ee
where we again limit ourselves to the contribution of carbon and oxygen as primaries, and we take $b_{CLi}\approx 0.12$ and $b_{OxLi}\approx 0.08$. As discussed in the previous section, equations (\ref{eq:reacSec}) and (\ref{eq:lithium}) can be solved by iterations, although it is not the only way. 

\section{Comparison with AMS-02 data}
\label{sec:data}

The spectra of primary nuclei (carbon and oxygen in our case) are calculated using equation (\ref{eq:iterapr}) but neglecting the role of reacceleration, for the reasons discussed in \S \ref{sec:primary}. The diffusion coefficient is assumed, as usual, to be only function of rigidity $R=p/Z=(A/Z)\sqrt{E_{k}^{2}+2 m E_{k}}$, where $Z$ is the charge of the nucleus and to have the following functional shape:
%\begin{equation}
\[
D(R)=
\left\{
\begin{aligned}
D_{0}~ & {\rm if\ R <R_{0}}\\
D_{0}\left( \frac{R}{R_{0}}\right)^{\delta_{1}} & {\rm if\ R_{0}\leq R\leq R_{1}}\\
D_{0}\left( \frac{R_{1}}{R_{0}}\right)^{\delta_{1}} \left( \frac{R}{R_{1}}\right)^{\delta_{2}} & {\rm if\ R\geq R_{1}}.
\end{aligned}
\right.
\]
%\end{equation}

%\begin{eqnarray}
%D_{0} ~~~ R<R_{0}\\
%D(R)=D_{0}\left( \frac{R}{R_{0}}\right)^{\delta_{1}} ~~~ R_{0}\leq R\leq R_{1}\\
%D_{0}\left( \frac{R_{1}}{R_{0}}\right)^{\delta_{1}} \left( \frac{R}{R_{1}}\right)^{\delta_{2}} ~~~ R\geq R_{1} .
%\end{eqnarray}

Throughout this section we consider several cases: 1) reacceleration with $\delta_{2}=1/3$; 2) Reacceleration with $\delta_{2}=1/2$; 3) Reacceleration with $\delta_{1}=\delta_{2}$; 4) No reacceleration and $\delta_{2}=1/3$; 5) No reacceleration and $\delta_{1}=\delta_{2}$. In all cases the value of $\delta_{1}$ and the normalization of the diffusion coefficient are obtained by comparison with the available data (both B/C and the spectral shape of C and O nuclei). 

The cases with $\delta_{1}\neq \delta_{2}$ are considered in order to mimic the spectral hardening observed in the spectra of primary nuclei, if such hardening at $\sim 300$ GV rigidity is in fact due to a change in the energy dependence of the diffusion coefficient at the same rigidity. One should keep in mind that in all physical models that describe the hardening the transition from the low to the high energy regime is gradual, while here for simplicity we assume that there is a sharp break in the diffusion coefficient. These toy models have the main objective of assessing the relative role of the reacceleration with respect to the presence of breaks in the rigidity dependence of the diffusion coefficient. We stress once more that we are especially interested in the range of kinetic energy per nucleon $E_{k}\gtrsim 10$ GeV/n, hence we may neglect both the effect of a possible advection with waves or winds and the effect of solar modulation. 

Numerically, in the description of the diffusion coefficient we choose $R_{0}=3$ GV and $R_{1}=336$ GV (in agreement with the recent AMS-02 fit to the proton and helium spectral breaks). 

\begin{figure}
\includegraphics[width=\columnwidth]{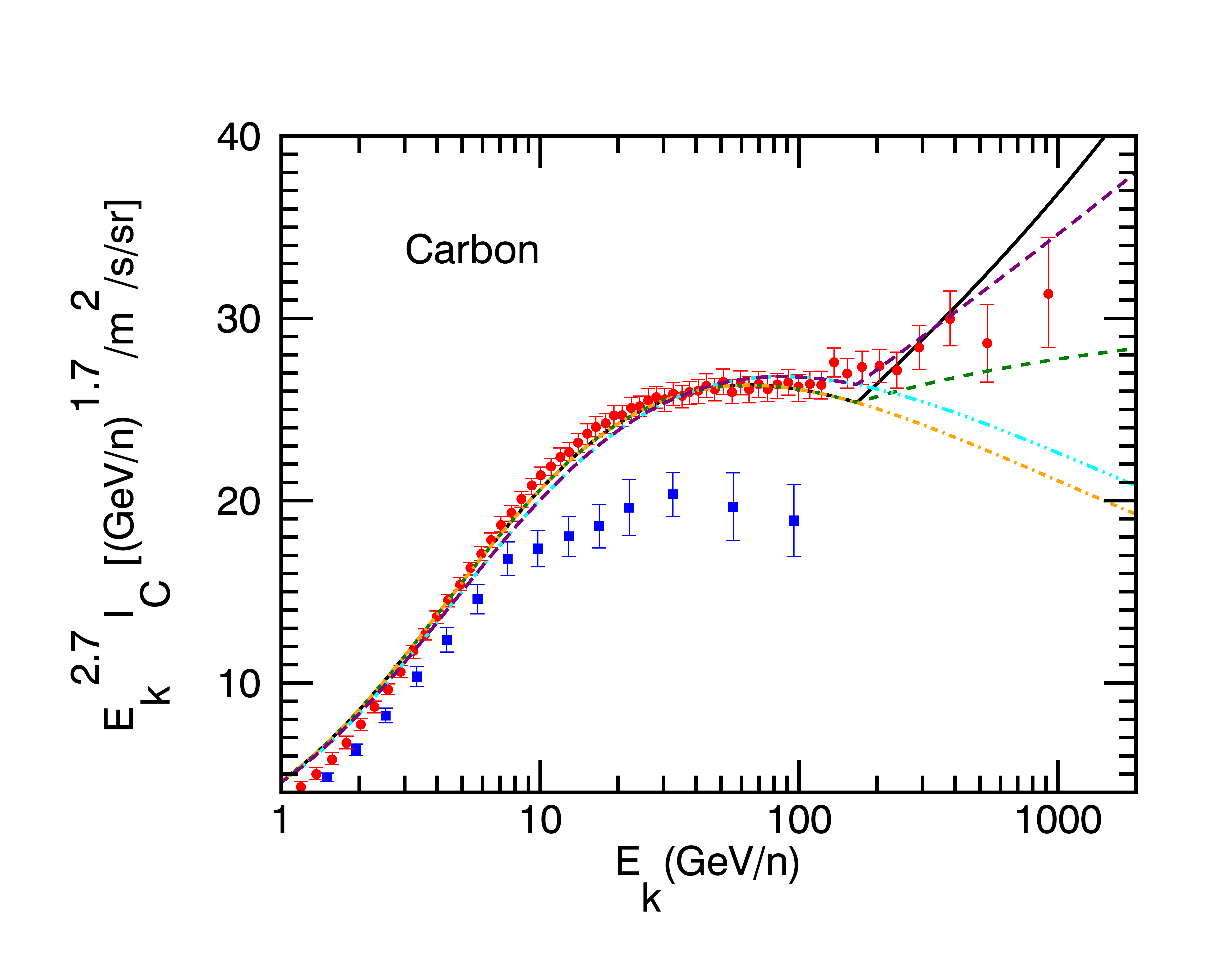}
\caption{Spectrum of Carbon nuclei: the red points are the preliminary results of measurements by AMS-02 \protect\citep{YanCern2017}, while the blue data points are the PAMELA data \protect\citep{2014ApJ...791...93A}. The lines illustrate our results for the following cases: 1) reacceleration with $\delta_{2}=1/3$ (solid black line); 2) Reacceleration with $\delta_{2}=1/2$ (dashed green line); 3) Reacceleration with $\delta_{1}=\delta_{2}$ (dash-dotted orange line); 4) No reacceleration and $\delta_{2}=1/3$ (dashed purple line); 5) No reacceleration and no break (dash-3dot cyan line).}
\label{fig:carbon}
\end{figure}

\begin{figure}
\includegraphics[width=\columnwidth]{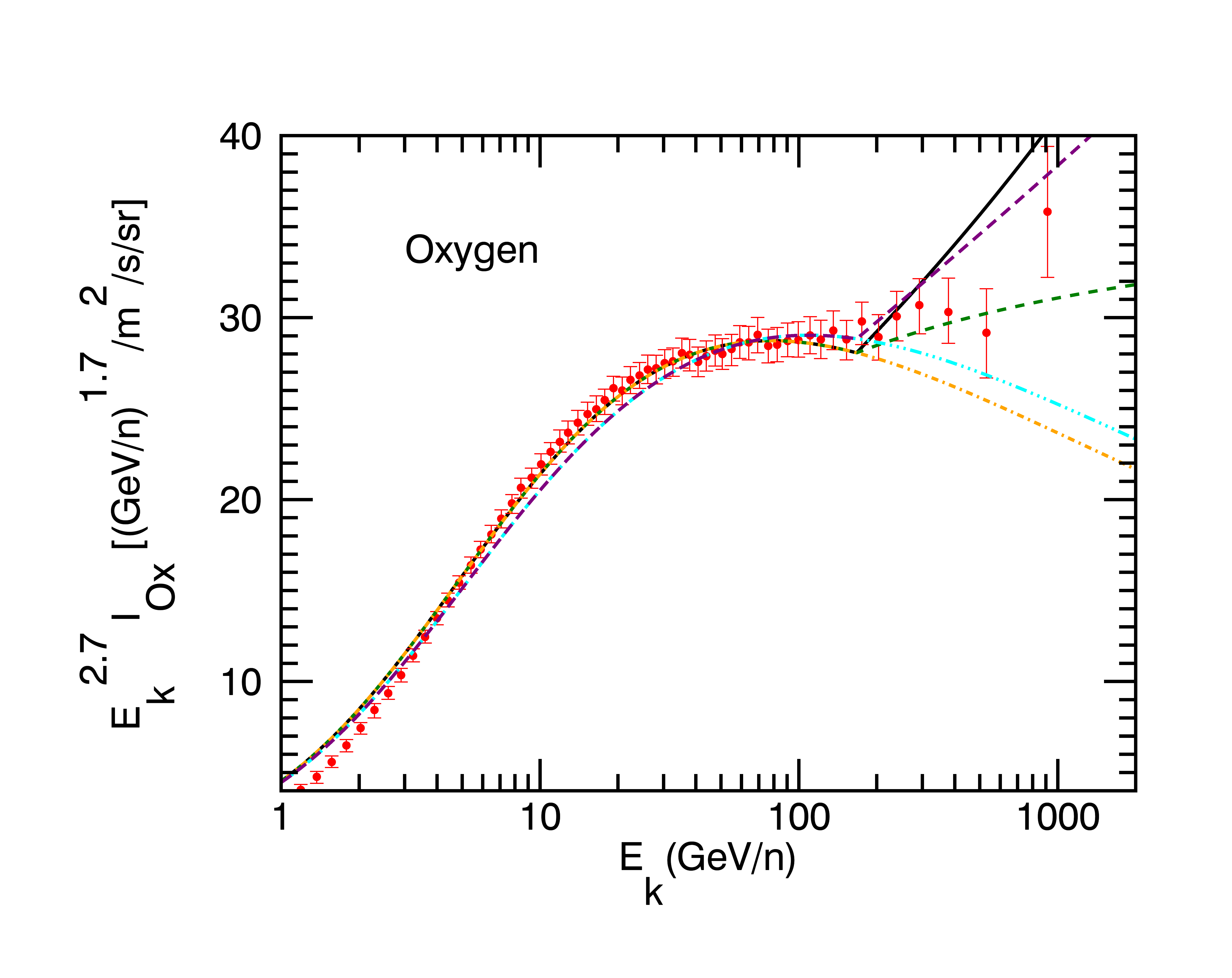}
\caption{Spectrum of Oxygen nuclei: the red points are the preliminary results of measurements by AMS-02 \protect\cite[]{YanCern2017}. The lines are labelled as in Figure \ref{fig:carbon}.}
\label{fig:oxygen}
\end{figure}

In Figure \ref{fig:carbon} and \ref{fig:oxygen} we show the spectrum of carbon and oxygen derived in our calculations in the five cases introduced above. As discussed earlier in this paper, reacceleration mainly affects the overall normalization of the spectra of primary nuclei, hence the calculation of these spectra is only used here as a way to normalize the relative spectra of C and O and as a check that the adopted grammage leads to no contradiction (for instance excessive or too small spallation of such elements). The lack of agreement between the predicted and observed spectra at energies below $10$ GeV/n is not surprising since we have not applied any correction for solar modulation here. Moreover, as mentioned several times above, the spectrum of primaries at such low energies is likely to be affected by advection, which is not included in the present calculation only to allow for a simpler interpretation of the results. For particle rigidity $\leq 336$ GV we impose that $s+\delta_{1}=4.85$, which is the best fit found by AMS-02 to the proton spectrum below the break. The value of $\delta_{1}$ is chosen so as to fit the B/C and the spectra of primary nuclei in the different cases. In the absence of reacceleration, one obtains $s=4.26$ and $\delta_{1}=0.59$. When reacceleration is included in the calculation, the B/C ratio is best fit with $s=4.19$ ($\delta_{1}=0.66$). Notice that the spectrum of carbon and oxygen at rigidity below $R_{0}$ is harder than that of protons, as a result of spallation reactions on these nuclei.

\begin{figure}
\includegraphics[width=\columnwidth]{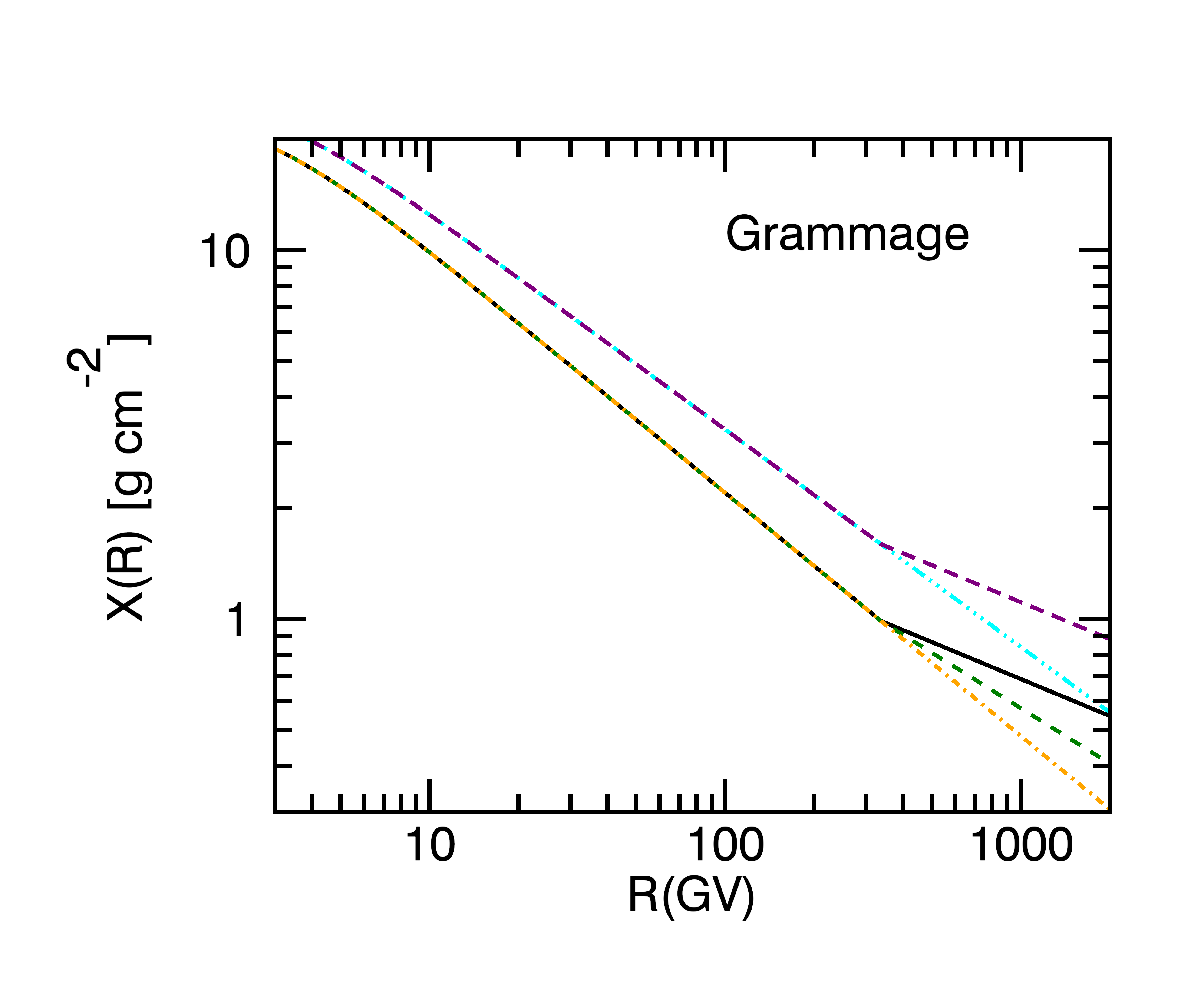}
\caption{Grammage traversed by CRs as a function of rigidity. The lines are labelled as in Figure \ref{fig:carbon}.}
\label{fig:grammage}
\end{figure}

The grammage resulting from the calculation reflects the normalization to the B/C ratio discussed below, and is shown in Figure \ref{fig:grammage} for the five cases listed above. It is worth noticing that reacceleration adds enough boron at low energies to require a lower grammage to fit the data. This also leads to requiring a steeper rigidity dependence of the diffusion coefficient when reacceleration is taken into account. 

The high energy behaviour of the spectra of primaries measured by AMS-02 (data points in figures \ref{fig:carbon} and \ref{fig:oxygen}) shows clear evidence for a hardening: for this reason we consider the cases in which $\delta_{2}=1/3$ and $0.5$. Although the latter seems to best describe the C and O high energy trend, one should keep in mind that a more realistic situation would show a gradual transition between the two regimes, hence we should probably not take these values too seriously but rather as phenomenological implementations of the idea of a transition in the diffusive properties at $R\sim 300$ GV.

\begin{figure}
\includegraphics[width=\columnwidth]{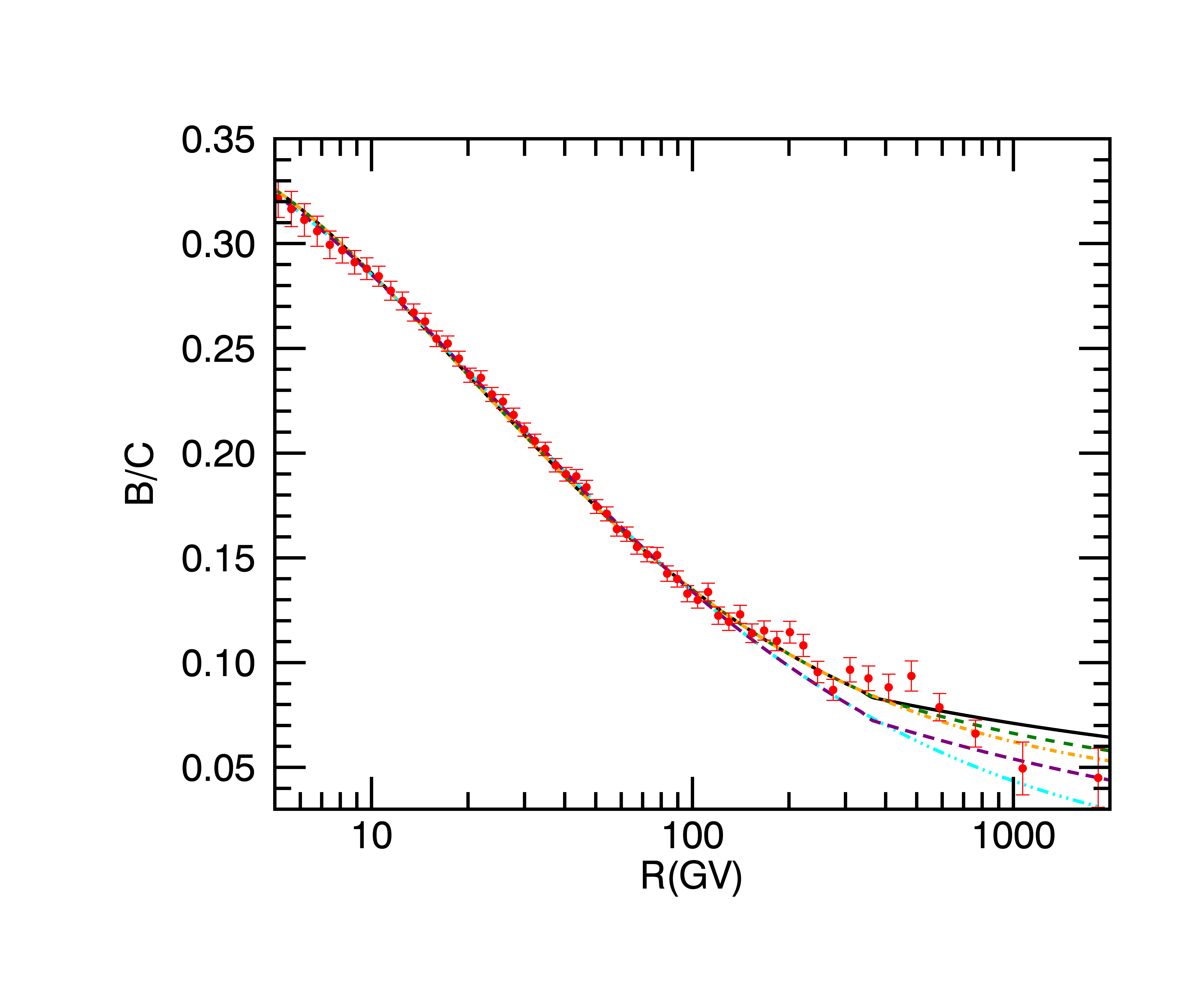}
\caption{Ratio of boron to carbon fluxes as a function of rigidity as measured by AMS-02 \protect\citep{2016PhRvL.117w1102A}. Lines are labelled as in Figure \ref{fig:carbon}.}
\label{fig:BC}
\end{figure}

The B/C ratio is much more interesting than the spectra of primary nuclei: in Figure \ref{fig:BC} we show the recent AMS-02 data \cite[]{2016PhRvL.117w1102A}, and the curves representing our predicted B/C ratio for the five cases introduced above. At rigidity $R<100$ GV, the five scenarios provide an equally good description of the data, although, as pointed out above, the grammage in the cases with and without reacceleration differ appreciably (figure \ref{fig:grammage}). However, for $R\gtrsim 100$ GV, the cases without reacceleration clearly fail to describe the B/C data points as measured by AMS-02. This was already pointed out by \cite{Aloisio:2015rsa} where the authors find that an additional grammage is necessary at high energies to fit the data, possibly accumulated inside the sources of CRs. Here we show that reacceleration at the same shocks responsible for CR acceleration may provide a better description of the B/C data, thereby mitigating the need for additional components to the grammage. Notice that reacceleration may occur at SN shocks even in the cases in which the shock propagates in a rarefied medium (for instance the ones excavated by the wind of the progenitor star) where no appreciable grammage is accumulated.

The dash-dotted (orange) line in Figure \ref{fig:BC} shows the predicted B/C ratio with reacceleration but without breaks in the diffusion coefficient. This case illustrates, by itself, the importance of reacceleration, and shows that reacceleration alone is sufficient to provide a good description of the observed B/C ratio. The introduction of a break in the diffusion coefficient ($\delta_{2}=1/3$ for the solid (black) line and $\delta_{2}=1/2$ for the dashed (green) line) leads to an additional flattening of the energy dependence in the B/C at high energy but does not lead to a clear improvement in the fit. 

The recent measurement of the spectra of secondary nuclei such as lithium and boron allows us to test further the ideas put forward above. In the absence of reacceleration, the spectrum of secondary nuclei is expected to scale with kinetic energy as $E_{k}^{-s+2-2\delta}$, where $\delta=\delta_{1}$ for rigidity below $\sim 300$GV and $\delta=\delta_{2}$ for rigidity $R\gtrsim 300$GV. The recent measurements carried out by the AMS-02 experiment show that the high energy spectrum of Li has a slope that is very close to that of the high energy spectrum of nuclei (slope $\sim 2.7$), apparently incompatible with the naive expectation based on the standard model.  

\begin{figure}
\includegraphics[width=\columnwidth]{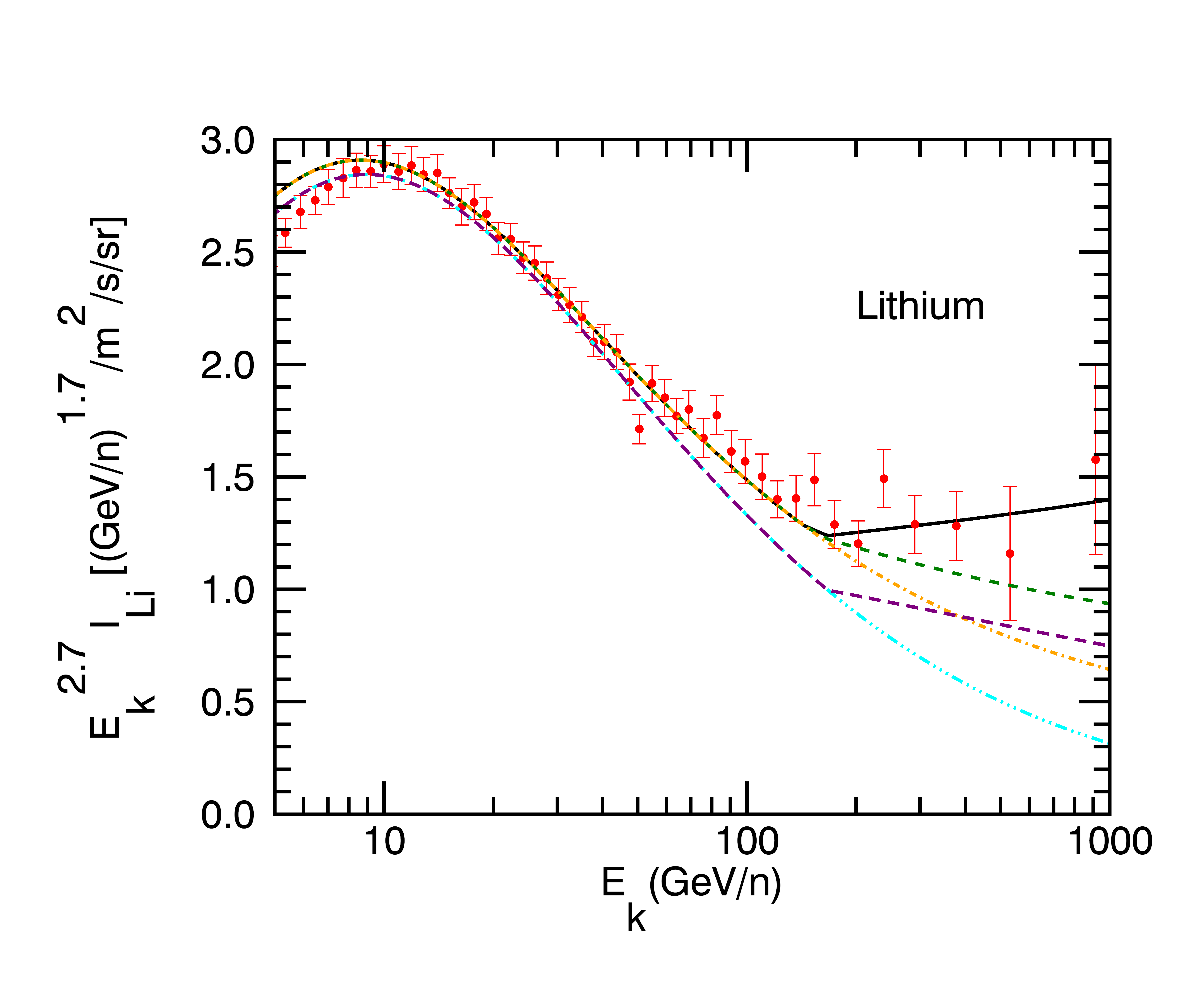}
\caption{Preliminary spectrum of lithium as measured by AMS-02 \protect\cite[]{YanCern2017} as a function of the energy per nucleon and results of our calculations (lines labelled as in Figure \ref{fig:carbon}).}
\label{fig:lithium}
\end{figure}

The spectrum of Li as measured by AMS-02 \cite[]{YanCern2017} is shown in Figure \ref{fig:lithium} and compared with the results of our calculations. The dash-3dot (cyan) and the dashed (purple) lines show the spectra of lithium in cases without reacceleration, without and with a break in the diffusion coefficient ($\delta_{2}=1/3$) respectively. One can see that these cases do not provide a good description of the observed Li spectrum. The presence of reacceleration drastically changes this picture: due to the fact that the spectrum of reaccelerated lithium nuclei (or any secondary nucleus for that matter) reproduces the spectrum of primaries at the same energy per nucleon, the high energy limit of such spectrum is the same as that of primaries because the steeper component disappears at lower energies. The data points of AMS-02 extend to a transition region between the low energy part, where the lithium spectrum scales as $E_{k}^{-s+2-2\delta_{1}}$ and the high energy limit where the spectrum is $E_{k}^{-s+2-\delta_{2}}$. The transition energy depends on how probable is for secondary lithium nuclei to encounter a SN shock, as can be understood by looking at equation (\ref{eq:lithium}). The curves shown in Figure \ref{fig:lithium} refer to the same cases that have been used to calculate the B/C ratio and to the spectra of C and O nuclei. Notice that the asymptotic limit in which lithium is dominated by reacceleration (slope $-s+2-\delta_{2}$) is reached only at energies $\gg 1$ TeV, not visible in the plot. 

The dash-dotted (orange) line shows that reacceleration alone (no break in the diffusion coefficient) is already sufficient to invalidate the naive expectation for the lithium spectrum at high energy. However the hardening in the lithium spectrum due to reacceleration alone seems to appear at too high energies to describe the preliminary data of AMS-02. On the other hand, adding the same spectral break that is necessary to describe the spectra of primaries (protons, He, C and O) one easily finds good agreement with the data, both for $\delta_{2}=1/3$ (solid black line) and $\delta_{2}=1/2$ (dashed green line) if reacceleration is taken into account. The break alone is not sufficient to explain the observed high energy lithium spectrum, as illustrated by the dashed (purple) line in Figure \ref{fig:lithium}. 

\begin{figure}
\includegraphics[width=\columnwidth]{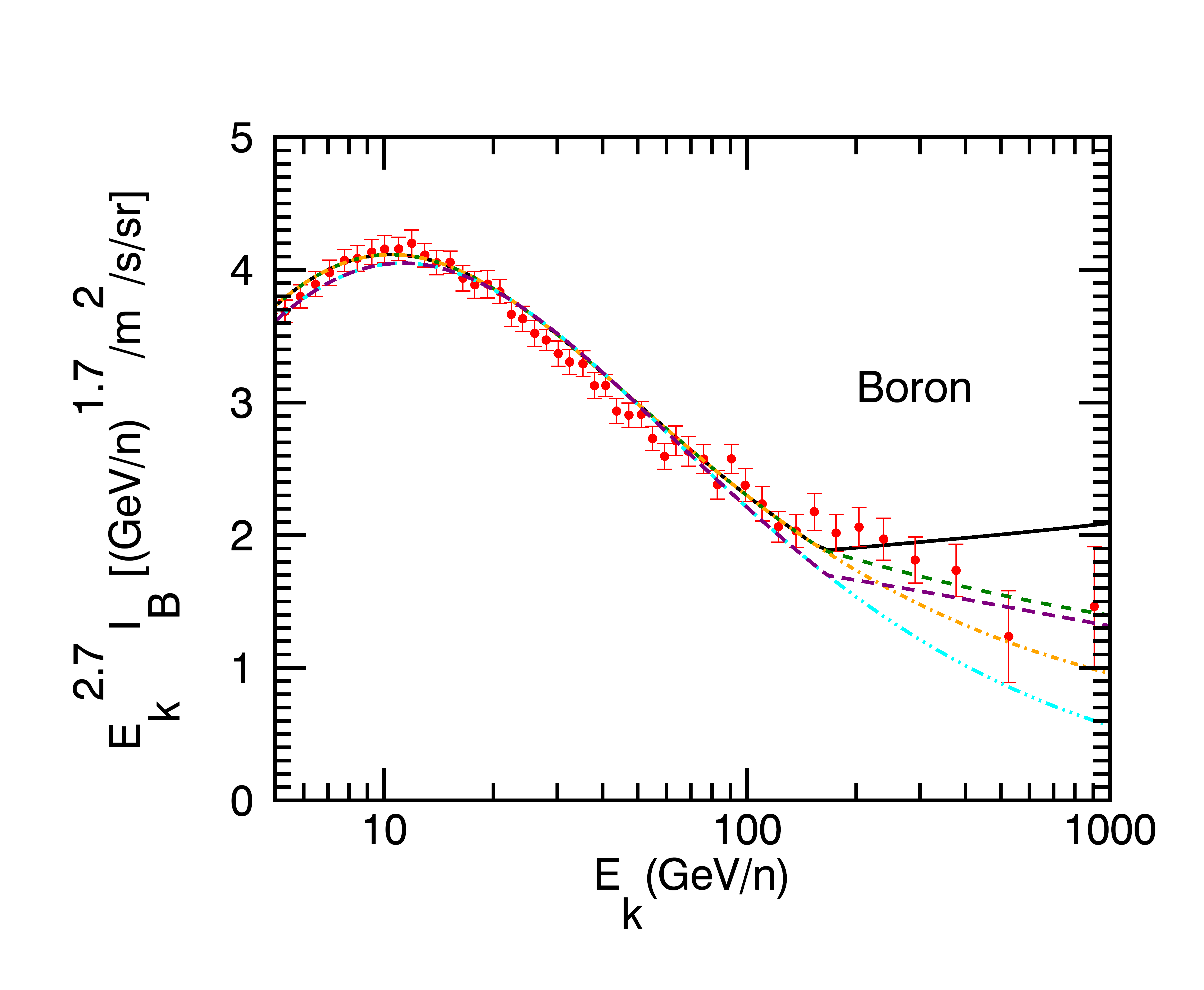}
\caption{Preliminary spectrum of boron as measured by AMS-02 \protect\cite[]{YanCern2017} as a function of the energy per nucleon and results of our calculations (lines labelled as in Figure \ref{fig:carbon}).}
\label{fig:boron}
\end{figure}

In Figure \ref{fig:boron} we also show the spectrum of boron nuclei, produced in the same reactions that give rise to lithium. One can see that qualitatively the same considerations already made for lithium hold for boron. While a hardening is visible in both spectra of boron and lithium at high energies, the level of such hardening seems to be mainly set, at present, by the last two data points at the highest energies, where statistical uncertainties are the largest. In this sense there is no indication of a significant discrepancy between the preliminary spectra of boron and lithium with what expected if they are pure secondary products. The presence of breaks and of shock reacceleration seems to account for the trend of both spectra.

Equation (\ref{eq:reacSec}) shows in a clear way that the effect of the reacceleration is strongly dependent upon the reacceleration volume, $\propto R_{SN}^{3}$, where $R_{SN}$ is the radius of a {\it typical} SNR. While all the results discussed above have been obtained by choosing $R_{SN}\sim 10-12$ pc, one might be tempted to speculate that by increasing the value of $R_{SN}$, the reacceleration term may be increased enough to explain the whole hardening of the lithium spectrum, with no apparent need for breaks in the diffusion coefficient. In fact this attempt typically fails, because the increase in the reacceleration term in equation (\ref{eq:reacSec}) (and equation \ref{eq:lithium}) also causes the low energy part to harden, resulting in secondary to primary ratios that do not fit the data. In addition, one should keep in mind that in the calculations illustrated above, the acceleration and reacceleration of particles at a SN shock have been modelled in a very simple manner, for instance ignoring all temporal evolution of the SN shock. This is important, because a different size of the shock corresponds to a different age of the remnant, which in turn corresponds to different maximum energies of the accelerated (and re-accelerated) particles: if $R_{SN}$ is too large, the assumption made implicitly in all calculations above, that the maximum rigidity is much larger than $\sim 1$ TeV/n may turn out to be inappropriate. For instance, if to assume that the maximum energy is regulated by the growth of Bell modes \cite[]{2004MNRAS.353..550B,2014MNRAS.437.2802S}, as in modern approaches to CR acceleration in SNRs, then one can write an approximate expression for the maximum energy (to be interpreted as rigidity for particles other than protons): 
\be
E_{M}(t) \approx \frac{\xi_{CR}}{10 \Lambda} \frac{\sqrt{4\pi \rho}}{c}e R_{SN}(t) v_{s}^{2}(t) \sim 100 ~ TeV \left( \frac{t}{t_{s}}\right)^{-4/5}.
\ee
valid for SNe in the Sedov phase (started at time $t_{s}$) exploding in the normal interstellar medium with a density of $\rho/m_{p}\sim 1~\rm cm^{-3}$. Here $\Lambda=\ln(E_{M}/(m_{p}c^{2}))\sim 10$ for a $E^{-2}$ spectrum, and $e$ is the proton electric charge. This maximum energy drops below TeV for times $\sim 100 t_{s}$, corresponding to radii $R_{SN}\sim 10-15$ pc (see \cite{2015APh....69....1C} for a more careful discussion on the derivation of the maximum energy). For SNe with a size bigger than $10-15$ pc it is likely that reacceleration only proceeds up to sub-TeV energies, even though the shock may still be strong. 

\section{Conclusions}
\label{sec:conclude}

The same shocks that are thought to be responsible for CR acceleration in the Galaxy are also bound to re-energize the ubiquitous CRs that happen to be in the region where the SN explosion takes place. While this effect certainly takes place, its strength is more uncertain, though we may expect it to be more important for older (bigger in size) SNRs than in the case of young (smaller) SNRs that CRs have a smaller probability to encounter in one escape time from the Galaxy. The spectrum of reaccelerated particles is the same as that of freshly accelerated particles: this simple consideration is sufficient to reach the conclusion that reacceleration is bound to be more important for secondary nuclei than for primaries. 

The recent precision measurement of the spectra of primary nuclei such as protons, helium, carbon and oxygen, of secondary to primary ratios such as the B/C ratio and of the spectra of secondaries such as lithium and boron stimulated a debate on whether such data confirm all the nuances of the so-called standard model of the origin of CRs or rather disprove it (see for instance \cite{2017PhRvD..95f3009L}). The B/C ratio measured by AMS-02 has been claimed to confirm that the diffusion coefficient increases with rigidity as $R^{1/3}$ \citep{2016PhRvL.117w1102A}, as expected for a Kolmogorov spectrum of turbulence. Such a scenario requires, at low energies, efficient second order Fermi acceleration in interstellar turbulence, so as to steepen the energy dependence of the B/C ratio and make it compatible with data. The plausibility of models with second order acceleration have recently been questioned by \cite{2014MNRAS.444..365D,2017A&A...597A.117D} based on the energy budget they require.

Moreover the recent detection of a hardening in the spectrum of virtually all elements in CRs led many authors to suggest that, for different reasons, the effective diffusion coefficient of CRs in the Galaxy may have a different energy dependence at low and high energies \cite[]{Tomassetti:2012ga,Blasi:2012yr}, the transition rigidity being $\sim 300$ GV. In this case the B/C ratio is also expected to change slope at the same rigidity.  

All these general considerations are based however on the standard paradigm in which the spectra of secondary nuclei are steeper than those of primaries by exactly $E^{\delta}$ where $\delta$ is the slope of the diffusion coefficient. This simple expectation fails when reacceleration at SN shocks is taken into account: at sufficiently high energy the spectra of secondary nuclei are dominated by the reaccelerated component, hence they replicate the shape of  the spectra of primary nuclei. For reasonable choices of the parameters this critical energy is well above TeV, hence, at the energies where current measurements are carried out one is always in a transition regime. We showed that this phenomenon can improve the description of the data on B/C as measured by AMS-02. In addition, reacceleration, together with the break in the diffusion coefficient required to explain the primary spectra, also accounts for the unexpected spectrum of lithium as measured by AMS-02 \cite[]{YanCern2017}.

The effect of reacceleration at SNR shocks is also expected to be important for antiprotons in that they also are secondary products of CR interactions in the Galaxy. The calculation of the flux of antiprotons is however more complex in that one needs to discriminate the effects of reacceleration and those associated with the energy dependence of the cross section for production of antiprotons. This investigation will be described in a forthcoming publication. Positrons are also expected to reflect the importance of reacceleration, but this phenomenon would lead, in the best case scenario, to a ratio $e^{+}/(e^{-}+e^{+})$ that tends to be constant as a function of energy. Hence the rising positron fraction that has been measured by both PAMELA \cite[]{2009Natur.458..607A} and AMS-02 \cite[]{2014PhRvL.113l1101A} still requires the existence of sources of freshly accelerated positrons. 

\section*{Acknowledgements}
The author is very grateful to the members of the Arcetri and GSSI groups for numerous discussions on the topic and for continuous, fruitful collaboration on related problems. E. Amato, C. Evoli, G. Morlino and S. Recchia were patient enough to actually read the manuscript and provide precious feedback. A special thank to R. Aloisio for help in collecting preliminary data of AMS-02.

\bibliographystyle{mnras}
\bibliography{boron} % if your bibtex file is called example.bib

\end{document}